# MetaTrinity: Enabling Fast Metagenomic Classification via Seed Counting and Edit Distance Approximation

**Arvid E. Gollwitzer[1,*], Mohammed Alser[2,*], Joel Bergtholdt[1], Joël Lindegger[1], Maximilian-David Rumpf[1], Can Firtina[1], Serghei Mangul[3], Onur Mutlu[1,*]**

[1]Department of Information Technology and Electrical Engineering, ETH Zürich, 8092 Zurich, Switzerland
[2]Department of Computer Science, Georgia State University, Atlanta, GA, USA
[3]Department of Clinical Pharmacy, University of Southern California, Los Angeles, CA, 90089, USA

*Corresponding author. Department of Information Technology and Electrical Engineering, ETH Zurich, Gloriastrasse 35, 8092 Zurich, Switzerland.
E-mail: arvidg@ethz.ch (A. E. G.), malser@gsu.edu (M.A.), omutlu@ethz.ch (O. M.)

## Abstract

Metagenomics, the study of genome sequences of diverse organisms cohabiting in a shared environment, has experienced significant advancements across various medical and biological fields. Metagenomic analysis is crucial, for instance, in clinical applications such as infectious disease screening and the diagnosis and early detection of diseases such as cancer. A key task in metagenomics is to determine the species present in a sample and their relative abundances. Currently, the field is dominated by either alignment-based tools, which offer high accuracy but are computationally expensive, or alignment-free tools, which are fast but lack the needed accuracy for many applications. To address this tradeoff, we introduce *MetaTrinity*, a novel metagenomic classification tool leveraging heuristic-based seed counting and edit distance approximation, achieving a balance of speed and accuracy that surpasses existing methods. We benchmark *MetaTrinity* against two leading metagenomic classifiers, each representing different ends of the performance-accuracy spectrum. On one end, Kraken2, a tool optimized for performance, shows modest accuracy yet a rapid runtime. The other end of the spectrum is governed by Metalign, a tool optimized for accuracy. Our evaluations show that *MetaTrinity* achieves an accuracy comparable to Metalign while gaining a 4x speedup without any loss in accuracy. Compared to Kraken2, *MetaTrinity* requires a 5x longer runtime yet delivers a 17x improvement in accuracy. This demonstrates a 3.4x enhancement in the accuracy-runtime tradeoff for *MetaTrinity*. This dual comparison positions *MetaTrinity* as a broadly applicable solution for metagenomic classification, combining advantages of both ends of the spectrum: speed and accuracy. *MetaTrinity* is publicly available at https://github.com/CMU-SAFARI/MetaTrinity.

# Introduction

Metagenomics extends traditional genomic analysis by studying the collective genome sequences of diverse organisms coexisting within a shared environment, enabling insights beyond single-species analysis. Comparing these complex datasets against large reference genome databases enables critical biological insights across fields such as microbial ecology, infectious disease monitoring, and early cancer detection.[1,2]. There are four key steps in a standard genome sequencing and analysis pipeline: 1) Genomic sequencing data is obtained through sequencing a new sample[3,4]. 2) Basecalling[5] procedures convert raw sequencing data into nucleotides A, C, G, and T in the DNA alphabet. 3) A quality control step[6] removes low-quality subsequences of a read or an entire read sequence. 4) Computational metagenomic analysis lists all taxa in the sample and their corresponding relative abundance levels[7].

Advancements in genome sequencing, driven by the successful completion of the Human Genome Project and the advent of high-throughput sequencing (HTS) technologies, have significantly accelerated research in clinical and life sciences. These technologies have not only expanded the scope of genomic analysis but also substantially reduced DNA sequencing costs, enabling broader adoption in medical diagnostics and microbial studies. Consequently, bioinformatics has developed many software tools to leverage increasingly large and complex sequencing read sets. These tools have triggered progress in modern biology and become essential to clinical life sciences. For instance, metagenomics has enabled advancements in precision medicine[1,8], understanding microbial diversity, and the early detection of diseases[9]. All this creates a need for faster and more efficient computational tools.

The increase in available metagenomic HTS datasets[10] has prompted the development of many taxonomic classification and abundance estimation methods, as evidenced by a recent benchmarking study involving a dataset established by the Critical Assessment of Metagenome Interpretation (CAMI) challenge[11]. CAMI covers 20 taxonomic classifiers, including both alignment-based approaches such as GATK[12], PathSeq[19] MetaPhlAn[13], and Metalign[14], and alignment-free approaches such as Kraken2[15], CLARK[16], KrakenUniq[17] and Centrifuge[18]. Early approaches for analyzing metagenomic sequencing data were alignment-based and used a reference database. However, the growth of HTS data and reference databases has made read search and alignment based on large databases computationally infeasible. On the other hand, alignment-free tools are less accurate than their alignment-based counterparts.

We consider a representative example detailed in the latest CAMI study: mOTUs[19], a highly accurate tool, requires four hours to perform taxonomic classification on the same marine (TARA Ocean[20]) dataset we consider in our analysis. This lengthy processing time renders it inapplicable for time-critical or routine procedures. Conversely, Kraken2[21] completes the task in less than six minutes but delivers a very low classification accuracy, with a high false-positive rate. Its F1 score[22] at the species level is only 0.03, an accuracy level that falls short for most medical applications[23].

Accelerating metagenomic analysis is critical for five key reasons:

1. The sequencing and basecalling steps for a sample read set are one-time tasks in most cases, while the reads from a single sequenced sample can be analyzed by multiple studies or at different times in the same study[24].
2. The analysis throughput is significantly lower than the throughput with which modern sequencing machines generate data. Sequencing throughput is expected to increase even further in the future.[25] As an example, Illumina NovaSeq X Plus systems[26], considering 10B Flow Cells, generate ~1 Tb of data per run (2 × 50 bp). This amounts to 20 billion reads (paired-end) passing filtering per flow cell. Our analysis, conducted with Metalign[14], a state-of-the-art metagenomic analysis tool, reveals that data analysis sequenced and basecalled by a high-throughput sequencer[12] in 48 hours lasts 38 days on a high-end server node. Such long analysis times present significant challenges for time-critical metagenomic use cases, including urgent clinical settings and outbreak tracing[27]
3. A single sequencing machine can concurrently sequence numerous samples from diverse sources, such as different patients or environments, thereby achieving remarkable throughput.
4. Furthermore, the extensive computational resources needed for metagenomic analysis make routine screening processes (i.e., for early cancer detection) practically inaccessible to the general public.
5. Sequencing technologies[28] that allow analysis during sequencing[29] further underscore the critical importance of fast metagenomic analysis.

We discuss the role of edit distance approximation methods in read mapping to quickly examine the similarity for every read sequence and potentially matching segments in the reference genome identified during seeding. Traditionally, the mapper performs computationally expensive sequence alignment to determine whether the remaining sequence pairs that pass the filter are similar. We observe that sequence alignment yields data not essential for metagenomic profiling, including the optimal number of edits, their precise locations, and the optimal arrangement of these edits[30,31]. These unnecessary computations waste compute cycles and energy. This necessitates performing the analysis on energy-intensive high-performance computing platforms that incur high costs and are unavailable in remote areas. Thus, many high-throughput applications like disease screening fall short of the possibilities enabled by metagenomic analysis.

We introduce MetaTrinity, a heuristic computational taxonomic classification approach based on Metalign[14]. Metalign is a state-of-the-art, highly accurate alignment-based tool for metagenomic analysis. We maintain the same level of accuracy as Metalign[14] but, through introducing heuristics, significantly enhance the speed of the analysis pipeline. Specifically, we:

1. Develop a memory-frugal reference database index structure that enables rapid reference database prefiltering, i.e., containment search. We achieve a 4x speedup over Metalign's reference database filtering procedure.
2. Accelerate the metagenomic read mapping phase by relying on heuristic methods for edit-distance approximation that provide close-to-optimal solutions significantly faster. Our heuristic and alignment-free read mapper delivers a close to fourfold runtime reduction benchmarked against minimap2.
3. Conduct a rigorous experimental evaluation to examine MetaTrinity's speed and accuracy. This involves comprehensively benchmarking our application against Metalign[14] to quantify our reduction in execution time.

# Results

We initially present a brief overview of the MetaTrinity algorithm. Then, we outline the organization of the reference database and index generation. We perform metagenomic analysis on simulated and real data sets to benchmark MetaTrinity against Metalign[14]. This evaluation includes accuracy metrics for the benchmarked tools and their computational resource usage.

## Methodology Overview

### The MetaTrinity Pipeline

MetaTrinity proceeds in three primary stages (Fig. 1). First, containment search (stage 1) identifies relevant reference genomes by counting how many seeds from the reads overlap with each genome. Only those genomes surpassing a user-defined seed-hit threshold are retained in a subset database, substantially shrinking the search space. Next, in heuristic read mapping (stage 2), each read is assigned to the most plausible locations within the filtered database using approximate edit distance metrics. We rely on established fast approximation techniques (e.g., SneakySnake, Hamming Distance, Base Counting) to rule out improbable matches. Finally, in taxonomic profiling (stage 3), we aggregate mapped reads to estimate species-level abundances, discarding taxa below a minimum abundance threshold.

The first stage (containment search) aims to filter the reference database by generating a much smaller *subset database* from the reference genomes that are likely to be similar to many reads in the metagenomic sample. To quickly quantify the similarity between a set of reference genomes and a set of reads, we perform seeding and count the number of seeds in each reference genome that are also present in the read set. This provides an estimate of how likely each reference genome will be present in the sample. The reference genomes with a number of seed hits above a certain threshold are then included in this subset database to reduce the unnecessary computations for analyzing highly dissimilar reference genomes.

In the second pipeline stage, we perform heuristic read mapping. Our read mapping stage aims to filter the metagenomic read set quickly and accurately in two steps. In the first read-filtering step, we examine the mapping locations of all reads and exclude candidate locations that do not achieve a minimum number of seed hits in the subset database. We exclude an entire read sequence if this read does not have at least one mapping location in the subset database with a number of seed hits above the threshold. After this step, we are left with a set of sequence pairs. Each sequence pair consists of a subsequence extracted from a specific mapping location in the subset database and a read sequence. The mapping regions in the subset database are determined based on the locations of seed hits. To accurately quantify the similarities for all the remaining sequence pairs, the second read filtering step uses heuristic algorithms to compute an approximate edit distance for each pair. Sequence pairs with an approximated edit distance above a user-defined cutoff threshold are filtered out from further analysis. We record the associated reference genome, mapping location, and edit distance for all remaining reads.

Based on the read mapping results, we perform taxonomic profiling and relative abundance estimation in the third stage. In this pipeline stage, we perform only one filtering step: we reduce the false positive rate and improve classification accuracy by excluding all organisms with relative abundance estimates below a user-defined cutoff threshold from the final taxonomic profile.

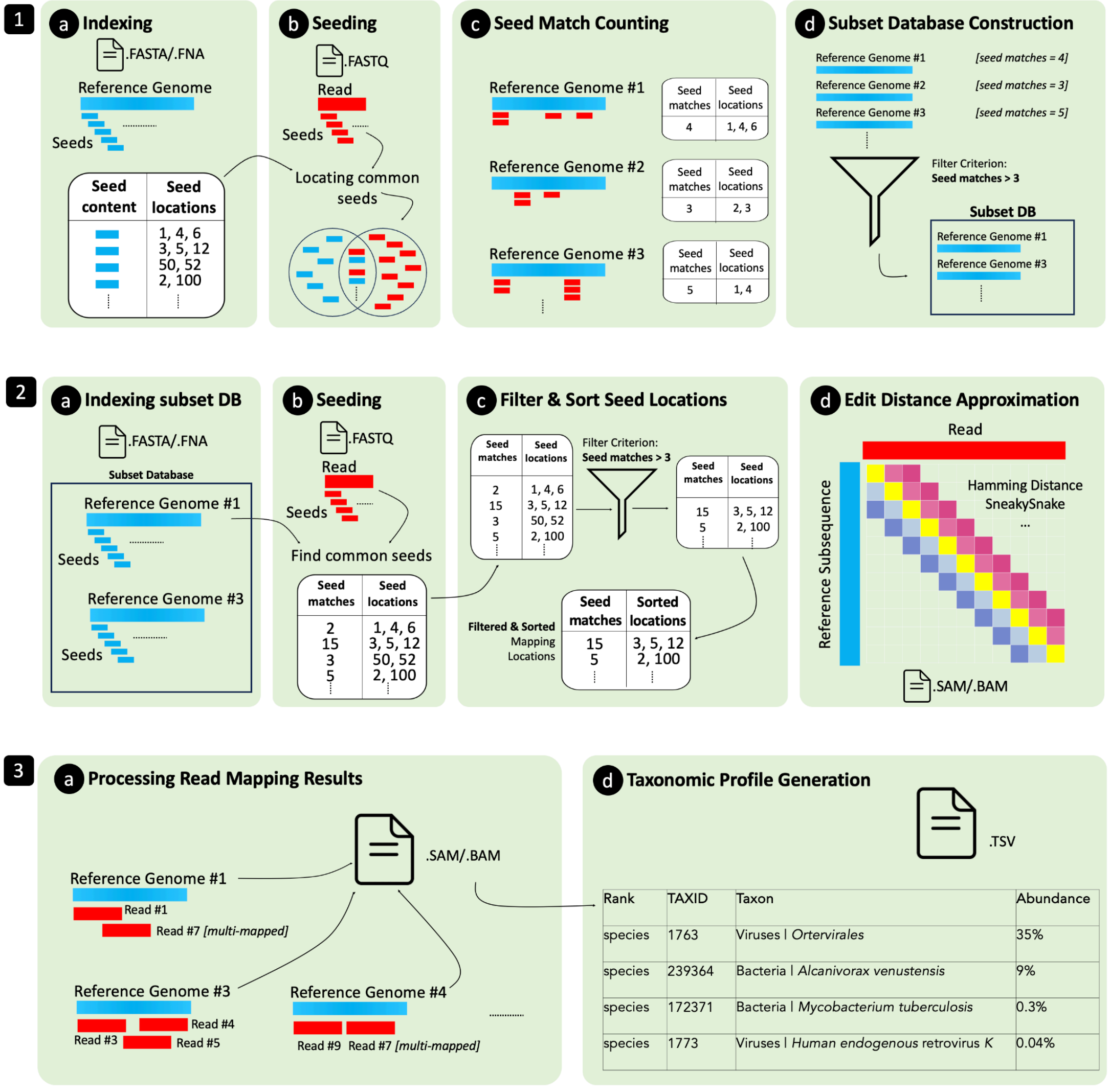

**Figure 1** Overview of the complete MetaTrinity pipeline.
**1)** MetaTrinity first uses Containment Search to build a small subset database. This database filtering stage begins with **a) Index querying:** MetaTrinity takes sequencing reads and a reference database as input. First, MetaTrinity collects seeds from all reference genomes in the database. **b) Seeding:** MetaTrinity extracts seeds from the input read set. **c) Seed Match Counting:** Our memory-frugal methodology quickly estimates the number of seeds in each reference genome that also exist in the reads. **d) Subset Database Construction:** MetaTrinity selects a significantly smaller subset database comprising reference genomes surpassing a certain seed hit count and builds a subset DB based on these genomes. **2)** MetaTrinity performs Heuristic Read Mapping, beginning with **a) Querying the subset database:** MetaTrinity loads the seeds from the small subset database created during containment search. **b)** We extract seeds from the read set and determine candidate mapping locations for each read by examining seeds simultaneously contained in the read set and the subset database, i.e., seed matches. **c) Filtering and sorting mapping locations:** In the first filtering step, MetaTrinity dramatically reduces the workload for later analysis steps by discarding all read candidate mapping locations that do not achieve a minimum number (in our case, three) of seed matches in the subset database. We then sort all remaining mapping locations by their associated seed matches. **d) Edit distance approximation:** MetaTrinity uses heuristics to approximately compute the edit distance for all remaining reads and mapping locations. We rely on several heuristic methods (like SneakySnake, Hamming Distance, SHD, etc.) to perform read mapping. All reads with a minimum edit distance exceeding 10% of the read length are discarded. **3)** We present the taxonomic profiling results in human-readable form by **a) Streamlined processing of mapping results:** MetaTrinity streams in the read mapping results as they are generated and computes the absence/presence and relative abundances of taxa on the fly. We consider reads uniquely mapped to a single reference genome, as well as multi-mapped reads. MetaTrinity examines the number of reads mapped to each reference genome and their respective edit distances to quantify relative abundance levels. **b) Taxonomic profile generation:** We discard all taxa with relative abundance levels below 0.01%. MetaTrinity then provides a taxonomic profile in the standardized format used by OPAL and CAMI as the final output.

## Containment Search

Containment search quantifies the similarity between a genomic dataset and a reference database by calculating the intersection of k-mers (short, fixed-length substrings of genomic sequences). This technique efficiently reduces the search space by identifying reference genomes with significant overlap in k-mer composition, making it particularly effective for rapid filtering prior to alignment-based methods. Containment search commonly involves building a containment index for a set of genomes to determine the intersection of k-mers between some dataset and the index.

Metalign[14] employs KMC3[32] and CMash[33] to select relevant reference genomes from a large database. CMash is a state-of-the-art hashing-based approach and uses a k-mer-based ternary search tree[34] (TST) to store variable k-mer sizes. Initially, KMC3 enumerates the k-mers in the reads and intersects these sets with the precomputed k-mers of the reference genomes. The containment MinHash similarity metric, i.e., CMash, then estimates the similarity or containment index between each reference genome and the input sample. The containment index, closely related to the Jaccard index[35], refers to the share of k-mers in a reference genome that also exists in the reads. Metalign includes all reference genomes above a specific cutoff threshold in a new, reduced database for alignment. We observe that KMC3+CMash generate more than four times the size of the examined reference database as auxiliary data during index construction. Furthermore, during index querying, KMC3 produces auxiliary data of roughly the size of the input read set, and CMash shows large main memory requirements. We propose a memory frugal indexing and efficient seed match counting algorithm to replace both KMC3 and CMash.

## Evaluation Methodology

We begin our analysis by examining the first two pipeline stages individually. We evaluate the computational resources of each stage. Specifically, we record the runtime and peak resident set size (RSS) for each phase: the first stage performs containment search and reference database filtering, while the second stage performs heuristic read mapping. Finally, we reintegrate the pipeline and analyze the entire system's accuracy. In the final step, we gauge the computational resources again, providing a comprehensive benchmarking strategy against Metalign. We benchmark our containment search algorithm, directly comparing it with KMC3+CMash. We use 64 compute threads for each tool on a system with an AMD EPYC 7742 64-core processor, 1 TB of main memory, and an SSD[36] with a SATA3 interface. We record the elapsed wall clock time and the main memory footprint in all experiments using the `/usr/bin/time -v` command on Linux. To ensure standardization, we use default parameters for KMC3 and CMash. The runtime measurements consider both the querying process and the generation of the subset database. Since the reference database filtering procedure employed by Metalign comprises two tools, namely KMC3+CMash, we perform a runtime breakdown to highlight each tool's individual contribution. We record a peak main memory usage of 14 GB for KMC3, independent of the read set. Later, when considering the complete pipeline, we continue our evaluation, focusing on accuracy.

To construct a comprehensive reference database, we incorporate NCBI[37] microbial genome assemblies, encompassing complete and incomplete assemblies from RefSeq[38] and GenBank[39]. The final database comprises 19,807 organisms, amounting to a size of approximately 170 GB. This large reference database forms the foundation for our genomic analysis. To enhance manageability and efficiency, we divide this database into N = 25 batches. We generate an index structure for each batch (MMI file). This offers numerous advantages, such as parallelized index access, which, as we will show, allows for considerable speedup. Each index (i.e., each MMI file) can be processed individually and independently in a multithreaded fashion. Moreover, our batch-based approach provides an efficient solution for handling updates to reference genomes. If a reference genome is updated, we only need to recreate the index for the affected batch. This localized update approach saves time and computational resources by eliminating the need to process the entire database. We then need to rerun MetaTrinity only for the affected batch. We count the seed hits for each reference genome in each batch. The number of compute threads to process an individual batch can be freely chosen. This step quantifies the similarities and differences between the reference genomes and the metagenomic read set. We include all reference genomes that receive a number of seed hits (*seed-hit count*) above an empirically determined threshold in the subset database.

To provide the most comprehensive analysis possible, we examine three datasets from the Critical Assessment of Metagenome Interpretation (CAMI)[11], one of each diversity class: a low-diversity sample with 99,796,358 reads (`RL_S001insert_270.fq`), medium-diversity with 99,776,814 reads (`RM_S001insert_270.fq`) and a high-diversity community (`RH_S001insert_270.fq`) with 99,811,870 reads. All CAMI datasets have a read length of 150. The CAMI paper[11] further details these communities. Of note, the CAMI communities include many organisms absent from the MetaTrinity reference database. For instance, for the CAMI high-complexity dataset, MetaTrinity's database only contains 184 out of the 243 unique species. We further underpin our accuracy and computational resource evaluation with a read set from the TARA Ocean Project[20] (`ERR1700889_1.fastq`), accessed February 2023. TARA Ocean reads have a length of 100 base pairs.

We must choose several parameters for our heuristic database filtering, i.e., the containment search approach. As we restrict our considerations to short reads, we rely on minimap2's short read parameters[40] (k-mer length $k = 28$; window size $w = 18$), which we assume to satisfy our optimality criteria. Since we are only interested in determining the number of seed hits for each reference genome, we construct all indices using minimap2's `-H` and `--idx-no-seq` options to speed up index construction, reduce storage space, and decrease the main memory footprint later during querying. At the end of our containment search stage, we include a reference genome in the subset database if it surpasses the normalized seed-hit count cutoff value.

Our goal is to find the minimum number of matching seeds that ensures a zero false negative rate. To achieve this, we iterate over several read sets (CAMI low, medium, and high complexity datasets and a TARA Ocean read sample) and incrementally decrease the minimum normalized seed hit cutoff value if a false negative occurs. For each read set, we divide all seed hits by the highest observed number of hits for normalization. We test the subset database for false negatives by performing read mapping with minimap2, then generate the taxonomic profile with Metalign's profiling routine and examine the false negative rate (on the species level) using OPAL[41]. Eventually, we make a conservative choice, selecting the lowest possible threshold that still ensures a zero false negative rate across all read sets. In this way, we determine an optimal threshold of 0.0001.

## Seed counting enables fast candidate genome identification

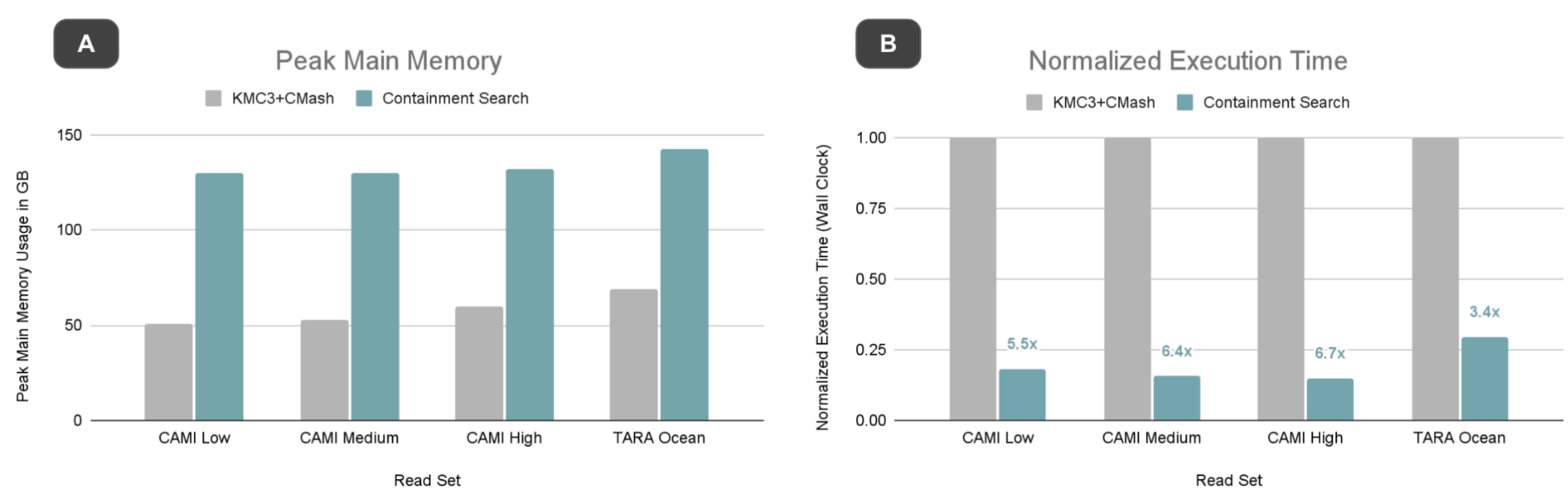

**Figure 2** Memory footprint and execution time analysis of our containment search stage. We benchmark our memory frugal containment search algorithm against the tools KMC3 and CMash employed by Metalign.
**A** Peak main memory usage in GB during the reference database querying and reference genome selection stage.
**B** Normalized execution time (wall clock time) of our containment search stage. We benchmark against KMC3+CMash.

We present the runtime analysis results in *Figure 2 A-B*. We make four key observations:

1. The runtime of KMC3+CMash remains relatively constant for the three CAMI challenge datasets and does not directly scale with the dataset size. The execution time does, however, decrease for the TARA Ocean dataset with a shorter read length of 100.
2. The runtime for our seeding algorithm increases with an increasing number of reads.
3. Our containment search algorithm achieves a 5.5x - 3.4x reduction in runtime.
4. Our containment search algorithm shows a two-fold increase in peak main memory usage. The peak main memory of our containment search approach remains relatively constant across all read sets.

The relatively constant execution time of KMC3+CMash for CAMI datasets is due to the generally rapid detection of a small overlap of k-mers from the read and reference by KMC3. As a result, even large read sets, which are highly dissimilar to most reference genomes, can be processed swiftly. However, this proves to be practically insignificant, given that the presence of specific organisms in the sample is unpredictable in advance. In contrast, our seeding algorithm iterates over all reads to count seed hits. Thus, its runtime is directly linked to the number of reads in the dataset. We conclude that our containment search algorithm generally provides significant speedup over KMC3 and CMash.

Our containment search methodology's increased memory footprint is related to the index size (in our case, the set of all MMI files). Given the same underlying set of reference genomes, our final indices show approximately twice the storage space requirements compared to the final index structures generated by KMC3+CMash. However, KMC3+CMash produces temporary auxiliary data during index construction, taking up 14 times more storage space than the final index structure. Our memory frugal methodology produces no temporary auxiliary files at all.In both cases, the complete index is loaded into the main memory during index querying. We can easily reduce the main memory footprint of our containment search stage by processing batches in the reference database sequentially, which, in turn, leads to an increase in execution time. Based on available hardware resources, our containment search methodology may be configured to minimize runtime or main memory usage by merely choosing appropriate user arguments.

We further analyze the index generation time of our seed-based, memory-frugal indexing methodology, comparing it against KMC+CMash on the same underlying reference data. In the initial step, we allocate 200 threads to each methodology. Our approach requires a five-minute execution time per batch, with a peak memory usage of 150 GB. Owing to the fully parallel execution, given the 200 available compute threads, the entire database is constructed within 5 minutes. On the other hand, KMC3+CMash requires 18 hours and a peak main memory of 300 GB for the same task. We repeat the procedure, this time limiting the thread count to 16 to accommodate realistic scenarios with limited hardware resources. Our method generates the index structure in 20 minutes with a memory footprint of 25 GB. We were unable to wait for the completion of the index structure generation by KMC3+CMash, which, given the now limited thread count, slowed down dramatically. We terminated the process after 22 hours. In this second attempt, the incomplete KMC3+CMash database generation needed a peak main memory of 700 GB.

## Alignment-free read-mapping

Locating potential subsequences within the reference genome sequence that bears similarity to the read sequence while accommodating differences remains a computationally intensive task. Many researchers aim to address this problem[42] by employing new algorithms, hardware accelerators[43], and hardware/software codesign[44]. Minimap2[40], our baseline, is a state-of-the-art read mapper that effectively maps nearly all existing sequencing read types, including short, ultralong, and accurate long reads. The operation of minimap2 covers four computational steps: 1) indexing, 2) seeding, 3) chaining, and 4) sequence alignment. Initially, minimap2 constructs an index database using minimizer seeds extracted from the reference genomes[31]. In the second step, the minimizer seeds extracted from a read sequence are matched to those extracted from the references. Third, the matching locations are sorted to identify adjacent seeds, which are then used to construct chains of matching seeds. Fourth and finally, a dynamic programming-based algorithm calculates sequence alignment between every two chains of seeds and stores mapping information in a sequence alignment/map (SAM and its compressed representation, BAM) file.

We propose accelerating minimap2 by refraining from the computationally expensive chaining and DP-based alignment algorithms. We observe that heuristics, in particular the sorting of mapping locations based on seed hits and edit distance approximation algorithms, achieve close-to-optimal results but at a much greater speed. This leads us to our new heuristic and alignment-free read mapper.

## Evaluation Methodology

We achieve read mapping using only heuristic methods. To that end, we perform seeding and evaluate the seed hits for each read. We record each read's mapping location and the corresponding number of seed hits. We exclude all mapping locations with less than three seeds. Subsequently, we sort the mapping locations by their associated seed hits in descending order. We then employ edit distance approximation algorithms to estimate the edit distance of each read for the first three mapping locations with the highest number of seed hits. In the following main filtering step, we exclude all reads that do not achieve an edit distance below a pre-defined threshold (usually 10% - 15% of the read length) for at least one mapping location. For all remaining reads, we record the mapping location and the associated estimated edit distance. We pass this information on to the final profiling stage and perform this procedure for all seven edit-distance estimation algorithms (all evaluated algorithms are detailed in Table 1).

In our selection of heuristic methods, we rely on the first comprehensive overview of edit distance approximation algorithms published from 1993 until 2020[45]. Edit distance approximation algorithms aim to estimate the edit distance between two sequences quickly. The sequences are dissimilar if the edit distance estimate surpasses a user-defined threshold. In genomic studies, sequences with an edit distance less than or equal to a user-defined threshold (E) are deemed biologically useful. All surveyed edit distance estimation approaches utilize an edit distance threshold to control rigor. The accuracy of edit distance approximation algorithms significantly depends on the edit distance threshold and the data analyzed. We employ thresholds of 10%, which we determine to be optimal in most real-world cases and data distributions (refer to supplementary materials). While technically feasible, higher thresholds are rarely biologically useful and thus are rarely observed in real-world applications.

We use one compute thread for each tool on the same system as before, i.e., an AMD EPYC 7742 64-Core Processor, 1 TB of main memory, and an SSD with SATA3 interface[36]. We benchmark our heuristic and alignment-free read mapper against minimap2. We again record the required computational resources, i.e., peak main memory and execution time. To obtain reliable runtime results, we perform each experiment three times and report the average of the observed execution times.

Our Analysis is again based on the three CAMI datasets, one of each diversity class: `RL_S001insert_270.fq`, `RM_S001insert_270.fq` and `RH_S001insert_270.fq` and a real-world read set from the TARA Ocean Project (`ERR1700889_1.fastq`). This read set was also considered in the latest CAMI challenge.

## Seed-based read filtering and edit distance approximation algorithms enable fast read mapping

We present our computational resource evaluation in *Figure 3 and Figure 4 A-D* and make five key observations:

1. Our alignment-free heuristic approach and minimap2 have practically the same main memory requirements.
2. Our alignment-free read mapper has a main memory footprint independent of the edit distance approximation algorithm. The peak main memory usage depends only on the read set and the reference database.
3. Methods such as Base Counting[46], Adjacency Filter[47], SneakySnake[48], HD[49], and SHD[50] consistently rank among the fastest. On average, these methods achieve a 3.5x speedup over minimap2.
4. In some cases, slower methods such as Magnet[51] yield only minimal speedup against minimap2.
5. If the workload for edit distance approximation algorithms is small, differences in runtime become barely visible, and all edit distance approximation algorithms deliver at least a 3x speedup over minimap2. We observe this effect for the CAMI low dataset.

**Figure 3** Comparison of the main memory footprint of our alignment-free heuristic read mapper against minimap2.

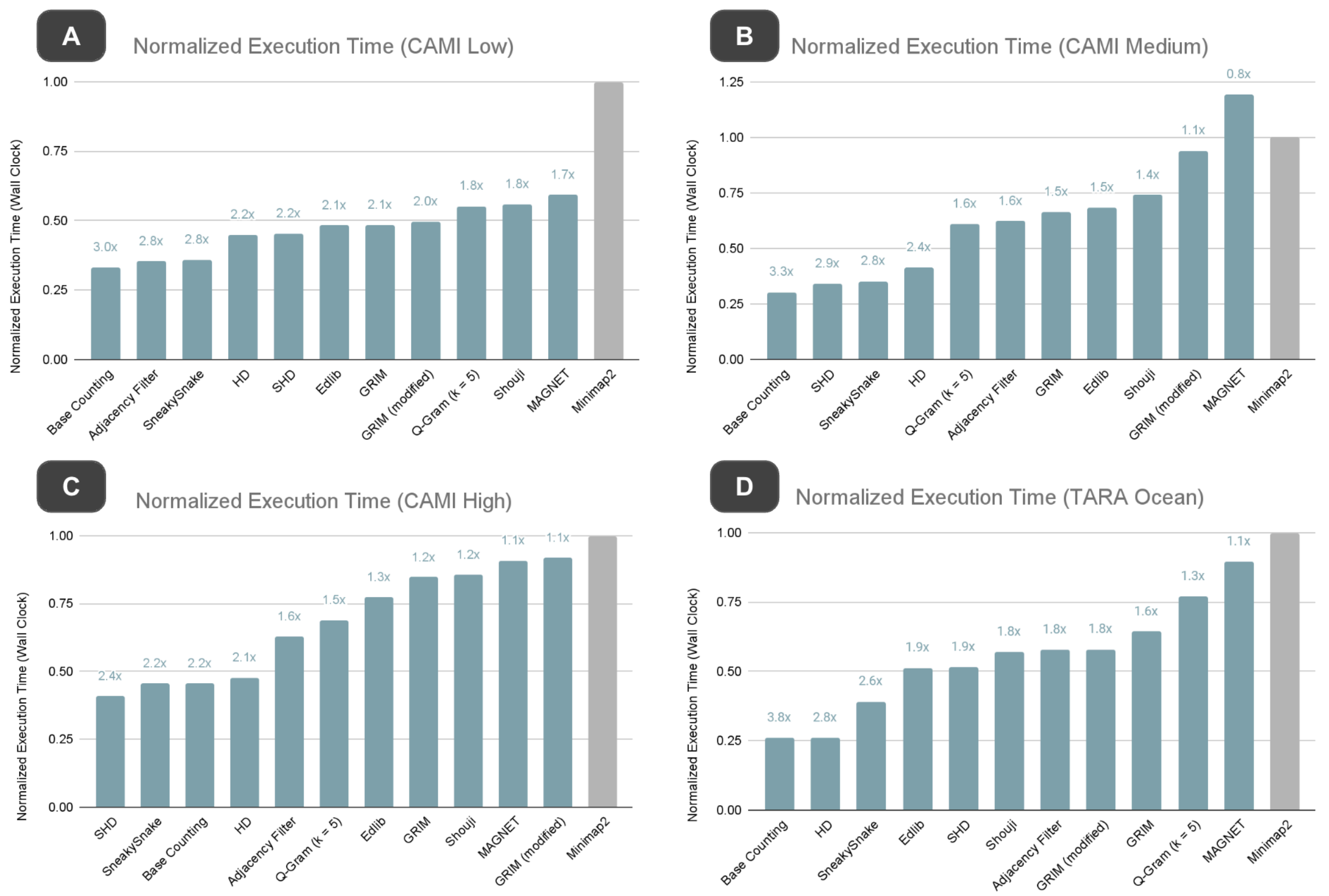

**Figure 4** Runtime analysis of our alignment-free and heuristic read mapper. We report the normalized elapsed wall clock time for each heuristic method employed in the read mapping stage and benchmark against minimap2. Each sub-figure shows the execution for each heuristic method and is normalized to the baseline runtime of minimap2.
**A** Normalized execution time is evaluated on the CAMI Low dataset. **B** Normalized execution time evaluated on the CAMI Medium dataset. **C** Normalized execution time evaluated on the CAMI High dataset. **D** Normalized execution time evaluated on the TARA Ocean dataset.

In conclusion, seed-based filtering and heuristic edit distance computation generally allow for strong speedup. The linear and some quadratic-complexity edit distance approximation algorithms (such as q-gram[52], GRIM[53], Base Counting[47], and SneakySnake[48]) deliver the greatest speedup. The differences in peak main memory usage stem from the index structure (MMI file), which is different for each read set, being loaded into main memory during index querying and seeding. The memory footprint is directly proportional to the size of the index, which in turn depends on the size of the subset database constructed in the containment search stage.

## The complete MetaTrinity Pipeline: Relative abundance estimation and genome identification

Efficient and accurate identification of each microbes' presence and relative abundances in an environmental sample directly recovered from its host environment continues to pose a significant challenge[54]. Some existing analysis techniques necessitate comparing the genomic composition of the subject sample to a large volume of genomic data and employing computationally intensive algorithms to identify a broad range of microbes[55]. This requirement confines the analysis to high-performance computing platforms, typically power-intensive and unavailable in remote areas. There remains a substantial need and room for improvement in existing metagenomic analysis tools[56,57].

We benchmark MetaTrinity against two leading metagenomic classifiers, each among the most dominant tools representing opposite ends of the performance-accuracy spectrum. On one end, Kraken2+Bracken[15], a tool optimized for performance, shows modest accuracy yet a short runtime. The high-accuracy end of the spectrum is governed by Metalign[14], a tool optimized for accuracy. Kraken2+Bracken relies on exact k-mer matching to assign taxonomic labels to all reads in a dataset. Bracken (Bayesian Re-estimation of Abundance with KrakEN) is a computational procedure for post-hoc estimating relative abundances. Bracken uses the classification outputs generated by Kraken2 and, through a Bayesian approach, refines the abundance estimates of taxa, leading to enhanced accuracy over the initial predictions offered by Kraken 2[58]. In general, k-mer-based techniques achieve rapid processing times but are known to deliver much lower classification accuracies than read-mappers like minimap2. Metalign[14], a state-of-the-art mapping-based metagenomic analysis tool, comprises three key steps. Initially, Metalign employs KMC3[32] and CMash[33] to narrow down the list of potential candidate organisms in the metagenomic sample. Secondly, Metalign uses minimap2[40] to map metagenomic reads to the filtered candidate genomes. Finally, Metalign estimates the relative abundances of microbes in the sample by amalgamating information from reads that uniquely map to one genome with those that align to multiple genomes. We propose our indexing and containment search algorithm combined with the heuristic read mapper presented earlier as the basis for a novel metagenomic classifier, MetaTrinity.

## Evaluation Methodology

We combine the previous two pipeline stages: 1) containment search and 2) alignment-free read mapping. Then we submit the read mapping results, i.e., reads mapped to one or multiple reference genomes and their associated edit distances, to the final stage for taxonomic profile generation. We benchmark MetaTrinity against Metalign and Kraken2+Bracken and measure peak main memory and end-to-end execution time. Our accuracy evaluation considers precision, recall, and the F1 score in genome identification, as well as the L1 norm error in relative abundance estimation. We evaluate the accuracy of MetaTrinity, Metalign, and Kraken2+Bracken using the CAMI-affiliated analysis software OPAL[41]. In our benchmarking strategy, we use datasets established by the Critical Assessment of Metagenome Interpretation (CAMI) challenge[11] and three real metagenomic datasets. Taxonomic ground-truth profiles provided by the CAMI[11] challenge serve as gold standards for our analysis on simulated datasets (*Figure 7 A-F*). We restrict our entire evaluation to the lowest taxonomic rank, i.e., the species level. MetaTrinity, Metalign, and Kraken2+Bracken have the same reference genomes in their respective reference databases. To draw robust conclusions on the accuracy of relative abundance estimation for each heuristic method in MetaTrinity's read mapping stage, we compute the average L1 norm error for each edit distance approximation algorithm for all read sets.

## MetaTrinity achieves fast and accurate taxonomic classification and abundance estimation

We thoroughly analyze the benefits of metagenomic classification at the species level with MetaTrinity.

We present our computational resource evaluation in *Figure 5* and *Figure 6 A-D*. We make three key observations:

1. MetaTrinity achieves a 4.5x speedup over Metalign for the methods SneakySnake, Base Counting, and Hamming Distance.
2. We generally observe an approximately 2-fold increase in peak main memory usage.
3. Hamming Distance, Base Counting, and SneakySnake remain the fastest methods.

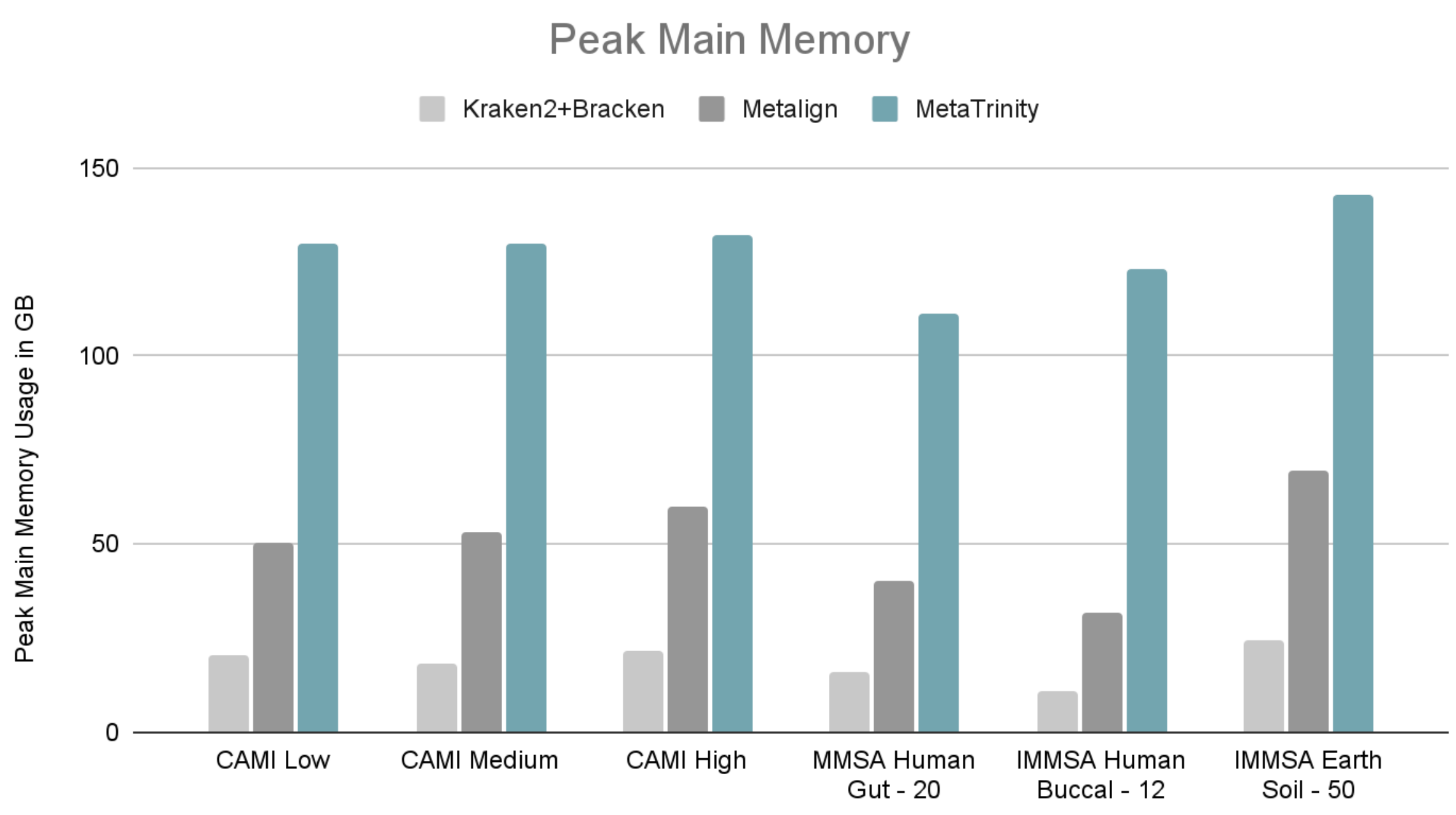

**Figure 5** Memory footprint analysis of the complete MetaTrinity pipeline. We benchmark against Metalign and Kraken2+Bracken.

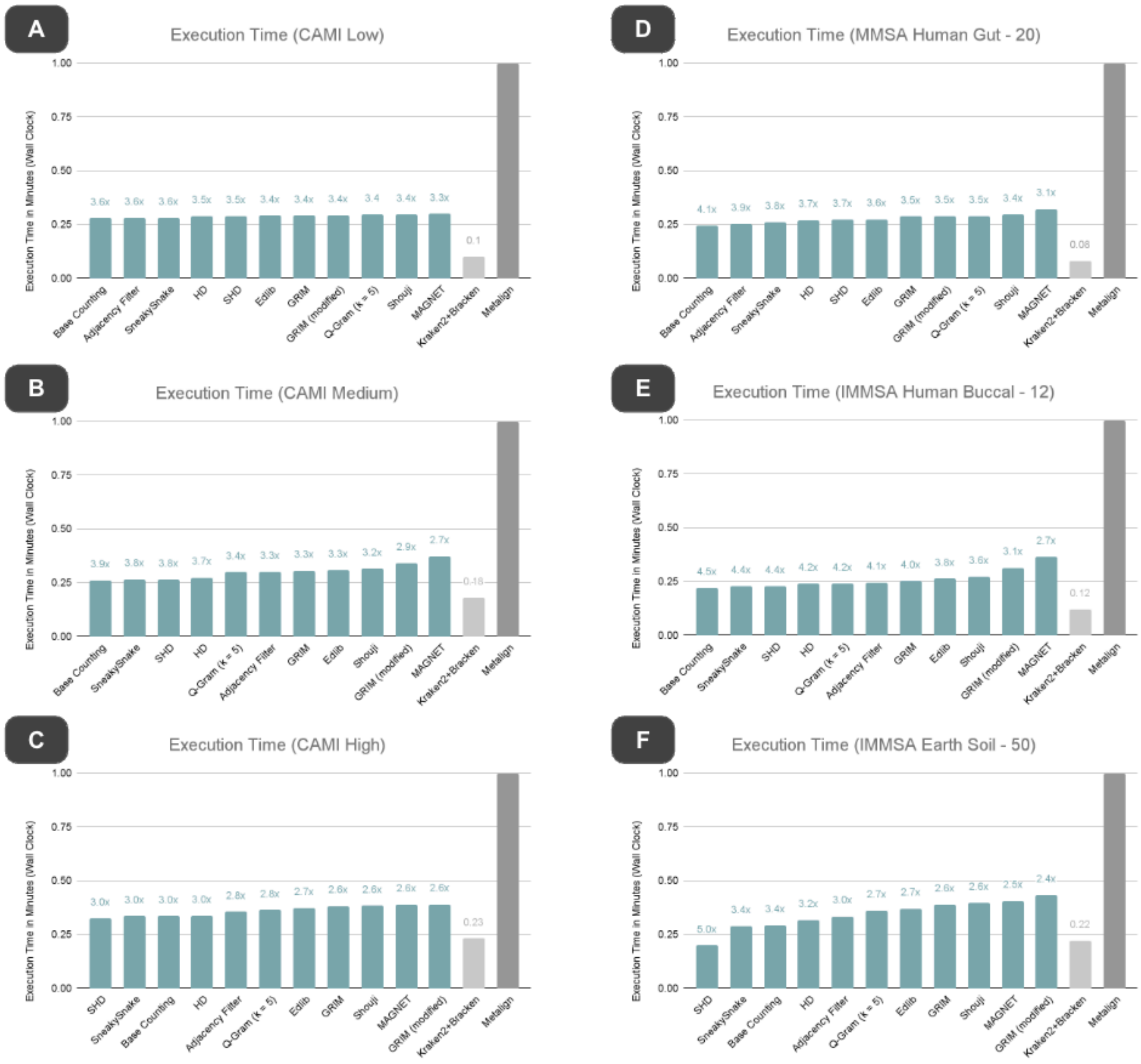

**Figure 6** Runtime analysis of the complete MetaTrinity pipeline on CAMI and IMMSA datasets. Benchmarked against Kraken2+Bracken and Metalign. We report the normalized elapsed wall clock time for each heuristic method employed in the read mapping stage and benchmark against Metalign. Each sub-figure shows the execution for each heuristic method and is normalized to the baseline runtime of Metalign. **A** Normalized execution time evaluated on the CAMI Low dataset. **B** Normalized execution time evaluated on the CAMI Medium dataset. **C** Normalized execution time evaluated on the CAMI High dataset. **D** Normalized execution time evaluated on the IMMSA Human Gut - 20 dataset. **E** Normalized execution time evaluated on the IMMSA Human Buccal - 12 dataset. **F** Normalized execution time evaluated on the IMMSA Human Soil - 50 dataset.

We conclude that MetaTrinity significantly reduces execution time for end-to-end metagenomic analyses. Since the metagenomic datasets we consider contain many species absent from MetaTrinity's database, the subset database constructed at the end of the containment search stage is relatively small (around 500 MB). As a result, the edit distance approximation algorithms employed in the read mapping stage are exposed to a lower workload, and the differences in execution time become less visible.

On our hardware (16 AMD EPYC 7742 64-Core Processors, 1 TB of main memory, an SSD with SATA3 interface[36]), the absolute execution time of MetaTrinity ranges from 32 min to 43 min, with most time spent on the containment search stage. Specifically, the containment search stage takes up approximately 16 - 18 minutes. The read mapping stage shows execution times ranging from 4 minutes (for the fastest edit distance approximation methods) to 14 minutes for the slowest edit distance approximation algorithms.

We present the classification accuracy results in *Figure 7 A-F, Figure 8 A-F* and *Figure 9*. We make three key observations:

1. All edit distance approximation algorithms exhibit zero false positive and zero false negative rates, resulting in an F1 score of always one.
2. Heuristic methods in the read mapping stage estimate relative abundance accurately, with minimal errors compared to the precise values computed by Metalign.
3. In the averaged L1 norm error analysis, Magnet[51], SneakySnake[48], and Edlib deliver the highest accuracy, i.e., the lowest relative abundance deviations

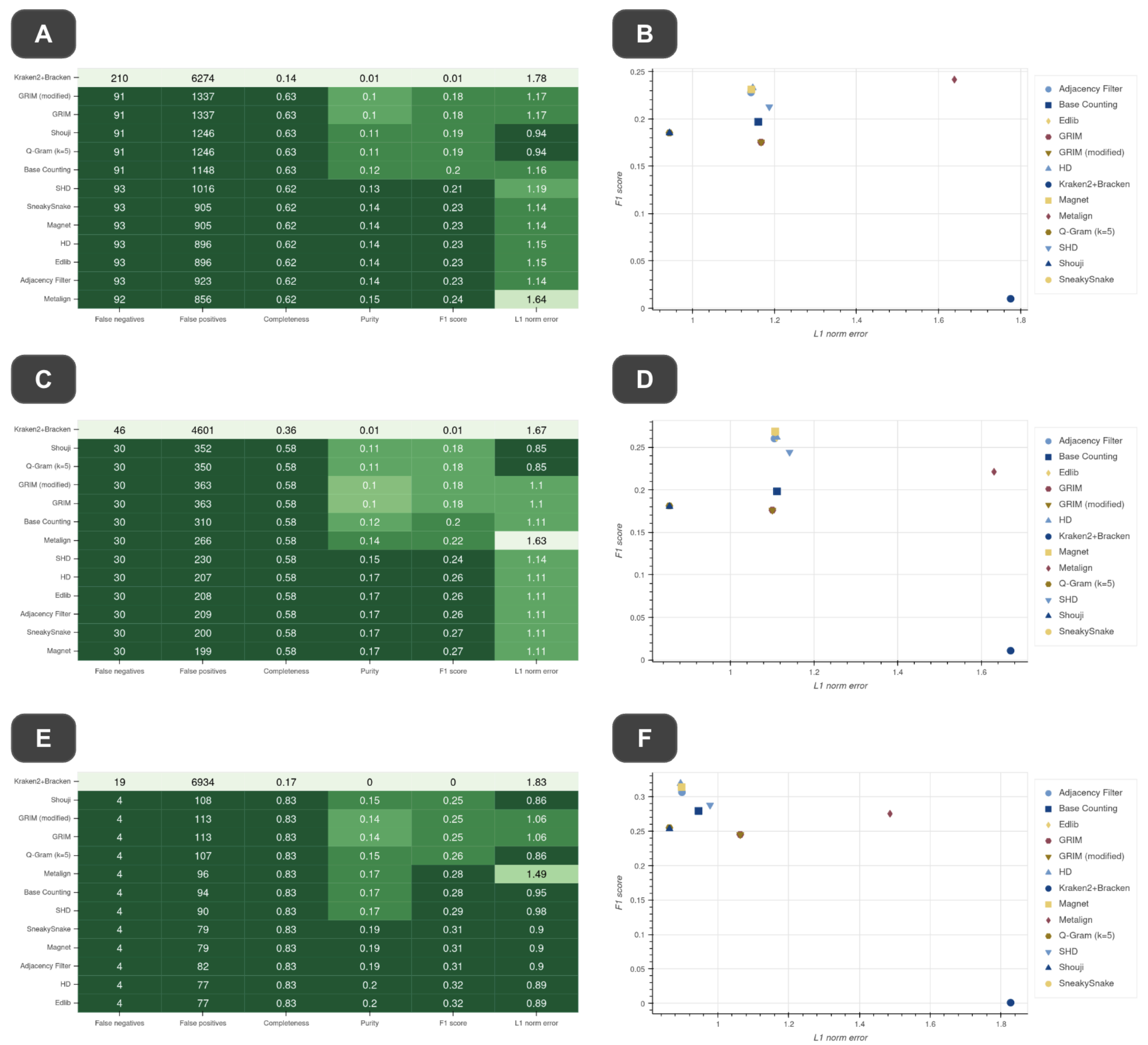

A

| | False negatives | False positives | Completeness | Purity | F1 score | L1 norm error |
|---|---|---|---|---|---|---|
| Kraken2+Bracken | 210 | 6274 | 0.14 | 0.01 | 0.01 | 1.78 |
| GRIM (modified) | 91 | 1337 | 0.63 | 0.1 | 0.18 | 1.17 |
| GRIM | 91 | 1337 | 0.63 | 0.1 | 0.18 | 1.17 |
| Shouji | 91 | 1246 | 0.63 | 0.11 | 0.19 | 0.94 |
| Q-Gram (k=5) | 91 | 1246 | 0.63 | 0.11 | 0.19 | 0.94 |
| Base Counting | 91 | 1148 | 0.63 | 0.12 | 0.2 | 1.16 |
| SHD | 93 | 1016 | 0.62 | 0.13 | 0.21 | 1.19 |
| SneakySnake | 93 | 905 | 0.62 | 0.14 | 0.23 | 1.14 |
| Magnet | 93 | 905 | 0.62 | 0.14 | 0.23 | 1.14 |
| HD | 93 | 896 | 0.62 | 0.14 | 0.23 | 1.15 |
| Edlib | 93 | 896 | 0.62 | 0.14 | 0.23 | 1.15 |
| Adjacency Filter | 93 | 923 | 0.62 | 0.14 | 0.23 | 1.14 |
| Metalign | 92 | 856 | 0.62 | 0.15 | 0.24 | 1.64 |

C

| | False negatives | False positives | Completeness | Purity | F1 score | L1 norm error |
|---|---|---|---|---|---|---|
| Kraken2+Bracken | 46 | 4601 | 0.36 | 0.01 | 0.01 | 1.67 |
| Shouji | 30 | 352 | 0.58 | 0.11 | 0.18 | 0.85 |
| Q-Gram (k=5) | 30 | 350 | 0.58 | 0.11 | 0.18 | 0.85 |
| GRIM (modified) | 30 | 363 | 0.58 | 0.1 | 0.18 | 1.1 |
| GRIM | 30 | 363 | 0.58 | 0.1 | 0.18 | 1.1 |
| Base Counting | 30 | 310 | 0.58 | 0.12 | 0.2 | 1.11 |
| Metalign | 30 | 266 | 0.58 | 0.14 | 0.22 | 1.63 |
| SHD | 30 | 230 | 0.58 | 0.15 | 0.24 | 1.14 |
| HD | 30 | 207 | 0.58 | 0.17 | 0.26 | 1.11 |
| Edlib | 30 | 208 | 0.58 | 0.17 | 0.26 | 1.11 |
| Adjacency Filter | 30 | 209 | 0.58 | 0.17 | 0.26 | 1.11 |
| SneakySnake | 30 | 200 | 0.58 | 0.17 | 0.27 | 1.11 |
| Magnet | 30 | 199 | 0.58 | 0.17 | 0.27 | 1.11 |

E

| | False negatives | False positives | Completeness | Purity | F1 score | L1 norm error |
|---|---|---|---|---|---|---|
| Kraken2+Bracken | 19 | 6934 | 0.17 | 0 | 0 | 1.83 |
| Shouji | 4 | 108 | 0.83 | 0.15 | 0.25 | 0.86 |
| GRIM (modified) | 4 | 113 | 0.83 | 0.14 | 0.25 | 1.06 |
| GRIM | 4 | 113 | 0.83 | 0.14 | 0.25 | 1.06 |
| Q-Gram (k=5) | 4 | 107 | 0.83 | 0.15 | 0.26 | 0.86 |
| Metalign | 4 | 96 | 0.83 | 0.17 | 0.28 | 1.49 |
| Base Counting | 4 | 94 | 0.83 | 0.17 | 0.28 | 0.95 |
| SHD | 4 | 90 | 0.83 | 0.17 | 0.29 | 0.98 |
| SneakySnake | 4 | 79 | 0.83 | 0.19 | 0.31 | 0.9 |
| Magnet | 4 | 79 | 0.83 | 0.19 | 0.31 | 0.9 |
| Adjacency Filter | 4 | 82 | 0.83 | 0.19 | 0.31 | 0.9 |
| HD | 4 | 77 | 0.83 | 0.2 | 0.32 | 0.89 |
| Edlib | 4 | 77 | 0.83 | 0.2 | 0.32 | 0.89 |

**Figure 7** Accuracy analysis of the complete MetaTrinity pipeline based on simulated CAMI Challenge datasets. We report the false negative and false positive rates, completeness, purity, and the L1 norm error for each heuristic method employed in the read mapping stage and benchmark against Metalign and Kraken2+Bracken. We consider all accuracies at the species level. **A** Overview of accuracy metrics for each heuristic method, evaluated on the CAMI Low dataset. **B** F1 score and L1 norm error for each heuristic method, evaluated on the CAMI Low dataset. **C** Overview of accuracy metrics for each heuristic method, evaluated on the CAMI Medium dataset. **D** F1 score and L1 norm error for each heuristic method evaluated on the CAMI Medium dataset. **E** Overview of accuracy metrics for each heuristic method evaluated on the CAMI High dataset. **F** F1 score and L1 norm error evaluated on the CAMI High dataset. **G** Overview of accuracy metrics for the TARA Ocean dataset. **H** F1 score and L1 norm error for the TARA Ocean dataset.

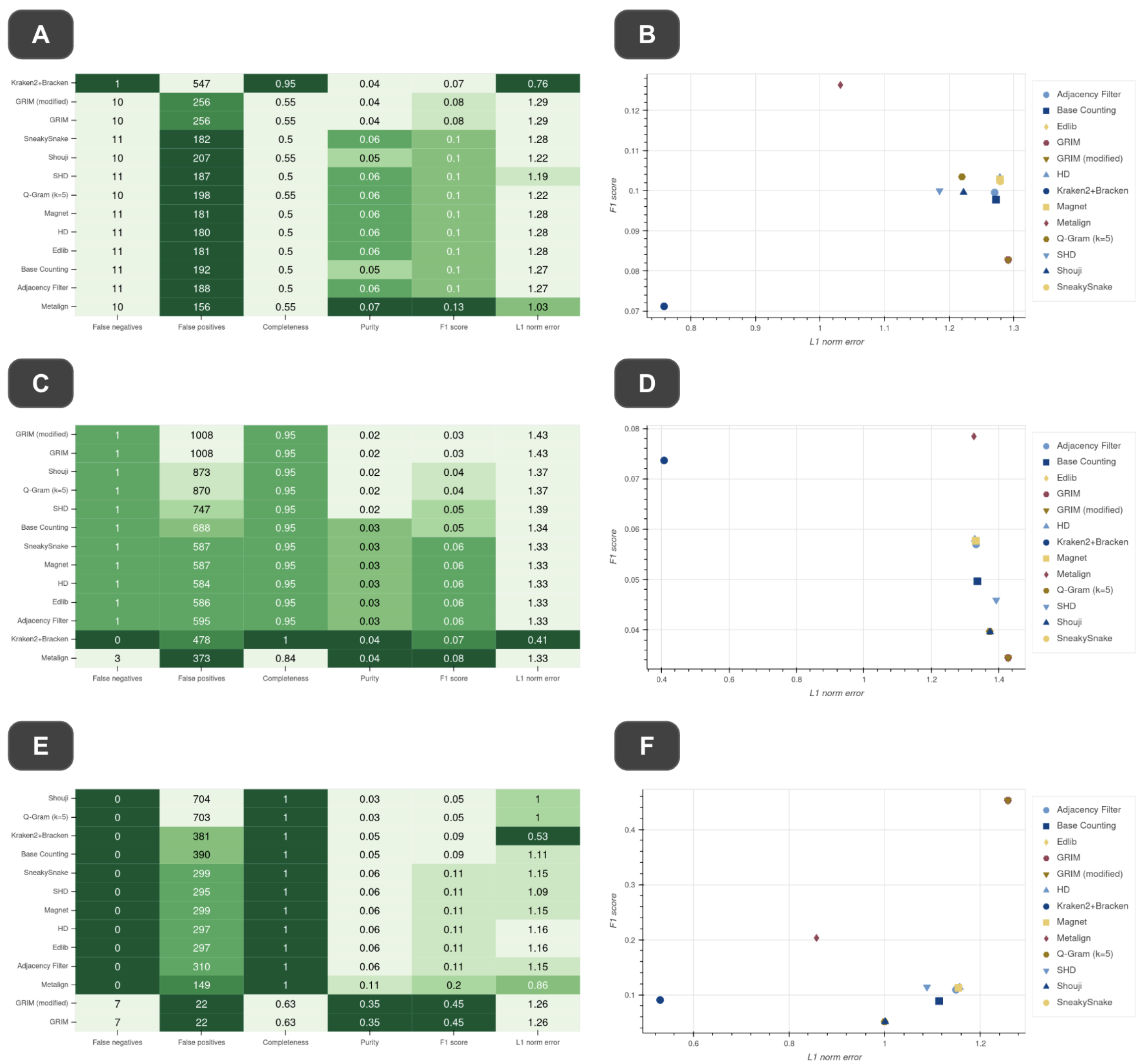

**Figure 8.** Accuracy analysis of the complete MetaTrinity pipeline based on real human microbiome datasets. We report the false negative and false positive rates, completeness, purity, and the L1 norm error for each heuristic method used in the read mapping stage and benchmark against Metalign and Kraken2+Bracken. We consider all accuracies at the species level. **A-B** Accuracy evaluation using human microbiome data from the Human Microbiome Project Database (SRX055381). **C-D** Accuracy analysis of taxonomic profiling accuracy on a human fecal sample (SRX13554335) from PRJNA747117. **E-F** Accuracy evaluation of MetaTrinity using the standardized NIBSC GutMix Rep 5 dataset (SRX8063904) from PRJNA622674.

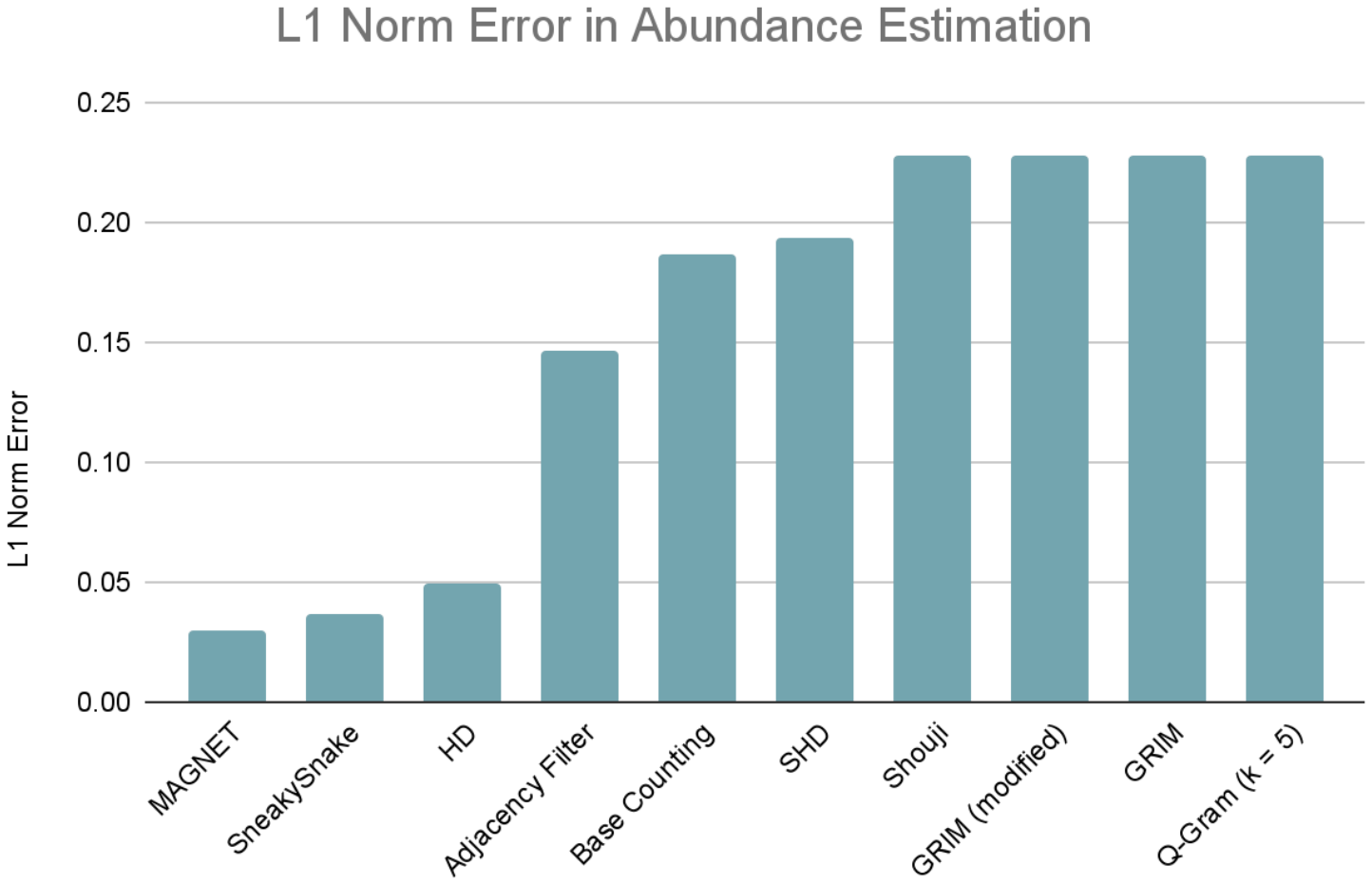

**Figure 9** L1 norm error for each heuristic method employed in the read mapping stage averaged over all four evaluated read sets.

A zero false negative rate in the final classification results again confirms a zero false negative rate in the reference database filtering stage. Upon combining the accuracy analysis with our previous computational resource evaluation, we conclude that SneakySnake offers the best accuracy-runtime tradeoff, while Base Counting stands as the fastest algorithm overall. Magnet is highly accurate in relative abundance estimation but too slow to gain significant speedup over minimap2, and one could simply use an aligner if the read mapping stage is required to deliver such high accuracy. The choice of edit distance approximation algorithm allows optimization for either runtime or accuracy: SneakySnake is the optimal choice for achieving the overall best accuracy-runtime tradeoff; Base Counting may be chosen to optimize for runtime.

In every taxonomic profiling experiment, we ascertain the correctness of each algorithm, or combination of algorithms, by confirming the presence or absence of each taxon with the correct results generated by Metalign. We consistently observe identical presence or absence of each taxon in all taxonomic profiles for each edit distance approximation algorithm. Consequently, the F1 score invariably equals one. Nevertheless, we notice minor discrepancies between the relative abundance estimates of Metalign (assuming the taxonomy profile of Metalign as the ground truth) and our edit distance approximation algorithms. These differences stem from the varying numbers of mapped reads provided by our algorithm and minimap2. We quantify these differences and represent them as L1 norm error, a measure for deviations in relative abundance. MetaTrinity correctly and accurately identifies the presence, absence, and relative abundance of taxa in a metagenomic sample.

# Discussion & Future Work

We introduced MetaTrinity, a computational tool designed to identify and quantify species abundance in a metagenomic high-throughput sequencing (HTS) sample. Our results demonstrate how tailoring alignment steps to the actual requirements of metagenomic classification can yield both higher throughput and robust taxonomic profiles. By reducing or eliminating dynamic programming in favor of fast approximate methods, we accelerate read mapping without compromising detection and abundance estimation. Many downstream genomic analyses similarly rely on partial or approximate read alignments, suggesting broader applicability of MetaTrinity's heuristic approach.

MetaTrinity enhances the entire metagenomic classification pipeline, reducing computational bottlenecks across all stages rather than optimizing isolated steps. This approach is crucial as focusing on a single stage limits the overall speedup, as Amdahl's Law[59] dictates. We note that improving read mapping performance impacts almost all genomic analyses that use sequencing data. For instance, read mapping constitutes up to 45% of the execution time in cancer genomics studies[60] and 30% in profiling the taxonomy of a multispecies sample[61]. Therefore, our accelerated read mapping stage may also prove useful as a standalone tool. In MetaTrinity, we leverage the similarities between indexing and seeding to develop a seeding-based containment search methodology. Indexing, which is off the critical path for most bioinformatics applications, is highly relevant for clinical and medical applications due to frequent modifications to the reference database.

In the reference database filtering, i.e., containment search stage, we observe a 6x speedup over KMC3+CMash when processing one batch, given that we consider reference databases based on the same reference genomes and the same read set. However, the speedup reduces to fourfold when we increase the number of batches and, correspondingly, the number of threads (assuming a thread count equal to or greater than the number of batches, i.e., at least one thread per batch). This slowdown arises from mutex waiting times and the necessity to combine individually generated results in a multithreaded scenario.

MetaTrinity's restriction to short reads is a potential limitation. The heuristic methods and empirically determined parameters are optimized for short reads, and extending them to long reads is not entirely trivial. However, there are abundant applications, for example, clinical metagenomics, where short-read sequencing technology prevails[62]. Sample collection, preparation, and sequencing are key steps in clinical applications[63]. Short-read sequencing technologies, such as Illumina[64], dominate clinical settings. Furthermore, the equivalent length of all reads stored in the same FASTQ file aids our use case, as most heuristic edit distance approximation methods spend most of their execution time on the longest sequences. The uniformity in length prevents a dramatic increase in runtime that could otherwise occur even in samples that contain only a few long sequences.

An optimization to be made in future works is reusing seeds generated from the reads in the containment search stage, later during read mapping. Currently, we perform seeding to locate seeds common to the read set and the full reference database to select a smaller subset database. Then, for read mapping, we repeat this seeding step with the same read set. Instead, we could simply use the seeds from our containment search stage. It still stands to be determined if we may also partially reuse the seed locations we found during containment search. However, since the subset database is very small (approximately 3 MB for CAMI High reads), index construction and seeding are fast. We never observed running times of more than two minutes for this repeated index generation and seeding phase. It further remains to be examined whether seeding in the read mapping stage significantly impacts runtime for very large read sets.

We anticipate that the foundation established in MetaTrinity will foster further advancements in metagenomic research. Specifically, we aspire for metagenomic-based screening and early diagnosis methods to become more prevalent and accessible in healthcare, drawing upon tools like MetaTrinity.

# Methods

The primary objective of MetaTrinity is a significant reduction in the end-to-end execution time of the indexing, seeding, and read mapping stages in metagenomic analyses. Given two genomic sequences, a reference sequence $R[0, ..., r - 1]$ and a query sequence $Q[0, ..., q - 1]$, where $r \geq$ q, these sequences consist of A, C, G, T in the DNA alphabet {A, C, G, T} in addition to the ambiguous base, N. Our goal is to locate all correct mapping locations of $Q$ in $R$ through a methodology that is fast, memory-efficient, and accurate.

## Reference Database Organisation and Index Construction & Querying

To compile a comprehensive reference database, we use all NCBI microbial genome assemblies[37], encompassing both complete and incomplete assemblies from RefSeq and GenBank, as of June 2020. The final database comprises 199,614 organisms, amounting to 170 GB in size (in gzipped form). For index construction, we divide the database into 25 batches, each containing approximately 7 GB of reference data, resulting in a 10 GB index structure (MMI) per batch. We never copy any reference genomes; the 7 GB reference data batch is merely a methodological construct. We employ minimap2-fast[65] to generate each batch's indices (MMIs) and store them in a separate directory. The batches are entirely independent during the index construction phase, with their interplay and possible combination occurring later during database querying. Consequently, indices for all batches can be generated in parallel, with the index generation time for a single batch being approximately five minutes. We refer to the entirety of all indices as our database.

During containment search, we query the database. An individual and independent minimap2-fast[65] instance can process each individual batch, i.e., each MMI file, either in a single or multithreaded fashion. We count the number of seed hits for each batch and then merge the results. Should a thread count lower than the number of batches be chosen (with at least one thread per batch) in the querying stage, we logically group several batches into clusters. We then process all indices in one cluster sequentially, with one thread. If a reference genome is simultaneously contained in several batches, we compute the number of seed hits for this reference genome for each batch. We then conservatively consider only the highest number of seed hits this reference genome achieves. We include all reference genomes that achieve normalized seed hits above the defined cutoff threshold in the subset database.

## Choice of Edit Distance Approximation Algorithm

There are four main edit distance approximation approaches for genomic sequence comparison. Based on prior benchmarking literature, we identify the methods with the most promising accuracy runtime tradeoff from each methodology.

**Table 1** Overview of the edit distance approximation methods surveyed.

| **Name** | **Year** | **Methodology** | **Short/ Long Reads** | **Native Platform** | **Language** | **URL Software** |
|---|---|---|---|---|---|---|
| SneakySnake[48] | 2019 | Pigeonhole | Short/ Long | CPU/ GPU/ FPGA | C/ C++ | https://github.com/CMU-SAFARI/SneakySnake |
| Shouji[66] | 2019 | Pigeonhole | Short | FPGA | C/ Verilog | https://github.com/CMU-SAFARI/Shouji |
| Hamming Distance (HD)[49] | 2019 | Pigeonhole | Short | CPU | | N/A |
| GRIM-Filter[53] | 2018 | q-gram | Short | PIM | C | https://github.com/CMU-SAFARI/GRIM |
| Magnet[51] | 2017 | Pigeonhole | Short | CPU | Matlab | https://github.com/BilkentCompGen/MAGNET |
| SHD[50] | 2015 | Pigeonhole | Short | SIMD | C/ SIMD | https://github.com/CMU-SAFARI/Shifted-Hamming-Distance |
| Adjacency Filter[47] | 2010 | Pigeonhole | Short | CPU | C | https://github.com/BilkentCompGen/mrfast |

Base Counting is the simplest method for comparing genomic sequences, as it merely compares the frequency of individual genomic bases between two sequences, resulting in a time complexity of $O(n)$[46]. The q-gram algorithm[52], seen as an extension of the Base Counting algorithm, compares the abundance of q-long subsequences to obtain a lower-bound estimate of the edit distance with an attractive runtime of $O(n)$. Finally, methods including SneakySnake and SHD are subject to the Pigeonhole Principle[67]. The simplest version of this principle is the Hamming Distance algorithm. Methods following the Pigeonhole Principle have a time complexity of $O(nE)$, where $n$ is the sequence length, and $E$ is the edit distance threshold, given as a percentage of the read length.

## Empirically Determined Parameters

We must choose several parameters for our heuristic database filtering and read mapping approaches. To achieve this, we sweep one parameter while holding all others constant. We repeat this in cyclic rotation to ensure that the choice of one parameter does not preclude the optimal choice of the others. This process continues until we find the optimal combination of parameters. We consider a parameter combination optimal if it leads to the highest accuracy-runtime tradeoff in the end-to-end accuracy and runtime evaluation of the MetaTrinity pipeline. In our containment search stage, we look for the optimal combination of two parameters:

- The seed length *k*
- And the minimum number of normalized seed hits a reference genome needs to achieve to be included in the subset database (i.e., the normalized seed-count cutoff value).

We iterate through all possible combinations of these two parameters and find that a seed length of $k = 28$ and a seed-count cutoff value of *0.0001* lead to the optimal end-to-end accuracy-runtime tradeoff.

For the alignment-free read mapping stage, we again aim to determine the optimal combination of two parameters:

- The minimum number of seeds per mapping location.
- The number of mapping locations with the highest number of seeds to examine.

We know the edit distance threshold to be optimally set to 10%[68].

Initially, we examine only the single mapping location with the highest number of seeds without requiring a minimum number of seed hits. We iterate over all four read sets as previously described. We profile the mapping results and examine the OPAL[41] report to identify species-level false negatives. If a false negative occurs for any read set, we include the mapping location with the next highest number of seeds. This inclusion continues until the false negatives are resolved.

Conversely, we first allow an unlimited number of mapping locations per read but require a minimum number of seed hits per read and mapping location. We start by considering all mapping locations with at least one seed hit. We incrementally increase the minimum number of seed hits until a false negative occurs in the taxonomic profile. Through this process, we determine the optimal choice to be a minimum number of three seeds per mapping location. All mapping locations not satisfying this criterion are discarded. A read that does not have at least one mapping location with at least three seed hits is filtered out entirely. We then consider the top three mapping locations per read, i.e., the locations with the three highest numbers of seed hits. If a read has fewer than three mapping locations with at least three seed hits, we consider the remaining locations that satisfy the minimum seed hits requirement.

## Taxonomic Profile Generation

In line with Metalign, we adopt the same parameters and choices, presuming their optimality for MetaTrinity. We observe a runtime of less than two minutes in the profiling stage for all considered read sets, a factor we deem negligible in our analysis.

## Optimization Strategies

We present two distinct optimization strategies for the containment search and read mapping stage, respectively.

### Performing aggressive seed-match based filtering

We can choose to increase the threshold of the minimum required number of seeds per mapping location. One increment already reduces the total number of reads for the subsequent edit distance approximation stage. This reduction in workload for the edit distance approximation methods leads to a significant speedup. This parameter can be set as a command-line argument. A high threshold, however, delivering great speedup, may cause false negatives in the final species-level taxonomic profile.

### Multithreaded containment search and parallelized index access

To maximize the parallel processing of our index structure, we initiate a thread for each batch, i.e., each MMI file in the reference database. We allow each file thread to launch ***t*** subthreads, thereby processing each MMI file in a multithreaded fashion with ***t*** parallel compute threads. We are interested in determining the seed-hit count for each reference genome. To achieve this, we keep track of the number of seed hits each reference genome receives through a data structure we call *seedmap*. Each seedmap uses the reference genome's accession number as key and stores the number of seed hits as value. We aim to efficiently balance the workload among all threads while minimizing waiting times arising from interdependencies of indifferent parallelly active compute threads. Our solution strategy begins by assigning each file thread a vector of seedmaps and a corresponding mutex for each seedmap. The number of seedmaps always equals the number of sub-threads. We do not limit our seed counting process (for one specific MMI file) to a single seedmap, as this would precipitate considerable waiting times for mutex availability.

We want to ensure an effective balance between the threads and the available seedmaps, thereby avoiding excessive mutex contention. To that end, we hash each accession number and then apply a modulo operation with the number of seedmaps (equivalent to the number of subthreads per MMI file). This relationship may be expressed as

$$\text{seedmap to access} = \text{hash(accession number)} \ \% \ \text{(number of seedmaps)}.$$

We increment the value associated with the accession number key in the corresponding seedmap by the seed count after securing the mutex. To construct the subset database, we must determine each reference genome's total seed-hit count. We thus merge all seedmaps, once containment mapping is concluded, by summing up the values of each key into a single seedmap. Given this final seedmap, we can now place all reference genomes with normalized seed-hit counts above the cutoff threshold in the subset database.

# Data & Code Availability

Throughout this paper, we exclusively rely on publicly available datasets. The CAMI challenge datasets (used in *Figure 7 A-F*) and ground truth profiles are available on the GigaDB website (https://doi.org/10.5524/100344).

For our evaluation on real data (*Figure 8 A-B*), we used human microbiome data from the Human Microbiome Project Database (https://www.hmpdacc.org/HMMC/). The sequencing data is available under accession SRX055381 (https://www.ncbi.nlm.nih.gov/sra/?term=SRX055381), with ground truth data derived from the OPAL repository (https://github.com/CAMI-challenge/OPAL/tree/master/data).

Our analysis in *Figure 8 C-D* was performed on a human fecal sample from PRJNA747117 (https://www.ncbi.nlm.nih.gov/bioproject/?term=PRJNA747117), with sequence data accessible under accession SRX13554335 (https://www.ncbi.nlm.nih.gov/sra/SRX13554335). Ground truth validation was based on the original study findings [69].

Our experiments in *Figure 8 E-F* used the NIBSC GutMix Rep 5 dataset from NCBI Bioproject PRJNA622674 (https://www.ncbi.nlm.nih.gov/bioproject/?term=PRJNA622674). The sequence reads are available under accession SRX8063904 (https://www.ncbi.nlm.nih.gov/sra/SRX8063904), with ground truth data sourced from the original study [70].

MetaTrinity's source code is available on GitHub: https://github.com/CMU-SAFARI/MetaTrinity.
For inquiries or if you wish to collaborate, please contact *arvid.gollwitzer@safari.ethz.ch*.

# Acknowledgments

The authors thank all members of the SAFARI Research Group for the scholarly environment they provide. We specifically thank Joel Bergtholdt for his contributions to this paper while pursuing an undergraduate research project under the mentorship of Arvid E. Gollwitzer.

Arvid E. Gollwitzer especially thanks Dr. Mohammed Alser and Prof. Onur Mutlu for their unwavering support and mentorship.